\documentclass[MPS,Times2COL]{WileyNJDv5}

\usepackage{algorithm}
\usepackage{algpseudocode}
\usepackage{bm}
\articletype{ORIGINAL RESEARCH}%

\received{Date Month Year}
\revised{Date Month Year}
\accepted{Date Month Year}
\journal{Journal}
\volume{00}
\copyyear{2023}
\startpage{1}

\begin{document}

\title{Optimized Pseudo--Linearization--Based Model Predictive Controller Design: Direct Data--Driven Approach}

\author[1]{Mikiya Sekine}
\author[2]{Satoshi Tsuruhara}
\author[3]{Kazuhisa Ito}

\authormark{M. Sekine \textsc{et al.}}
\titlemark{Optimized Design of Pseudo--linearization--based Model Predictive Controller: Direct Data--driven Approach}

\address[1]{\orgdiv{the Graduate School of Engineering and Science, Mechanical Engineering}, \orgname{Shibaura Institute of Technology}, \orgaddress{\state{307 Fukasaku, Minuma, Saitama 3378570}, \country{Japan}}}

\address[2]{\orgdiv{the Graduate School of Engineering and Science, Functional Control Systems}, \orgname{Shibaura Institute of Technology}, \orgaddress{\state{307 Fukasaku, Minuma, Saitama 3378570}, \country{Japan}} and also is \orgdiv{a research fellow with Japan Society for the Promotion of Science}, \orgaddress{\state{Tokyo 1020083}, \country{Japan}}}

\address[3]{\orgdiv{the Department of Machinery and Control Systems}, \orgname{Shibaura Institute of Technology}, \orgaddress{\state{307 Fukasaku, Minuma, Saitama 3378570}, \country{Japan}}}

\corres{Corresponding author Mikiya~Sekine. \email{md22050@shibaura-it.ac.jp}}

\presentaddress{307 Fukasaku, Minuma, Saitama 3378570, Japan}


\abstract[Abstract]{To reduce the typical time--consuming routines of plant modeling for model--based controller designs in SISO systems, the fictitious reference iterative tuning (FRIT) method has been proposed and proven to be effective in many applications.
However, it is generally difficult to properly select a reference model without a prior information on the plant. This significantly affects the control performance and might considerably degrade the system performance. To address this problem, we propose a pseudo--linearization (PL) method using FRIT, and design a new controller for SISO nonlinear systems by combining data--driven and model--based control methods. The proposed design considers input constraints using model predictive control. The effectiveness of the proposed method was evaluated based on several practical references using numerical simulations for hysteresis and dead zone classes and experiments involving artificial muscles with hysteresis characteristics.}


\jnlcitation{\cname{%
\author{Mikiya S.},
\author{Satoshi T.},
\author{K. Ito}}.
\ctitle{Optimized Design of Pseudo--linearization--based Model Predictive Controller: Direct Data--driven Approach}
\cvol{2021;00(00):1--18}.}

\maketitle

\renewcommand\thefootnote{}

\renewcommand\thefootnote{\fnsymbol{footnote}}
\setcounter{footnote}{1}

\section{Introduction}\label{sec1}

In the design of control systems for unknown industrial plants, two approaches are commonly used for tuning the controller parameters: trial and error, and applying mathematical model.
Both methods are time--consuming. The former requires numerous preliminary experiments, whereas the latter is highly complex as it requires model structure selection, identification, and evaluation to acquire a precise model\cite{ref1}.
Therefore, data--driven control methods were proposed\cite{ref2,ref3}.
These strategies use only input--output (I/O) data and do not require explicit or precise mathematical models.
Examples include the simultaneous perturbation stochastic approximation--based model--free control \cite{refSPSA}, virtual reference feedback tuning (VRFT) \cite{ref5}, fictitious reference iterative tuning (FRIT) \cite{ref6}, model--free adaptive control \cite{ref7}, and some other related methods \cite{ref4,ref8,ref9,ref10}.
In particular, FRIT and VRFT are direct controller--tuning methods.
These design controller parameters offline using only a single I/O data in a practical controller structure.
This significantly reduces the design effort required.
In addition, in FRIT, tuning is performed based on the output results, and depends on the initial reference \cite{ref11}.
Hence, an inverse reference model is not required.
Therefore, to select a proper transfer function that generates pseudo--signals for parameter tuning, a prefilter is unnecessary.
This is more practical when the reference before and after tuning remains the same.
%
However, while FRIT significantly reduces design efforts using single I/O data, it has some limitations.
FRIT--based methods have a drawback that they require a reference model in advance to generate the desired output for a closed--loop system.
Control performance deteriorates or becomes unstable if the reference model is inappropriately selected.
This is because these methods force the matching behaviors between a closed--loop system with an unknown plant, a designed controller, and a reference model.
To solve this problem, simultaneous tuning of the controller and reference model through optimization has been proposed \cite{E-FRIT}.
However, in this case, the direct controller--tuning method generates controller parameters that track the optimized desired output via a reference model, not the reference signal.
Consequently, the designer cannot tune the control performance.

To address this problem, we propose a new controller design \cite{IFAC}.
In the inner--loop, we introduced a pseudo--linearization (PL) model for model matching based on FRIT.
On the other hand, in the outer loop, we adopted model predictive control (MPC) as our model--based control (MBC) strategy.
This approach allowed us to consider the input differences and their constraints directly.
This integrated method improves tracking performance more closely for the reference signal rather than merely achieving the desired output via a reference model.
Furthermore, it allows for the tuning of MPC's design parameters such as cost function weights and prediction or control horizons, facilitating the reflection of the designer's intent.
In terms of MPC, the proposed method increases the efficiency of the control design, even for systems with unknown model structures.
Similar studies have proposed methods in which the reference model is decoupled from the system performance \cite{Bem,Wakitani}.
These studies applied MPC using a direct controller--tuning approach, as opposed to traditional mathematical modeling.
Notably, in one of these studies, data--driven control was applied to design an MPC considering the input constraints, and a Kalman filter was used to estimate the internal inputs \cite{Bem}.
Our proposed method further refines this concept using a PL model in conjunction with an optimized controller to estimate the internal inputs \cite{IFAC}.
Moreover, it enhances the control performance by optimizing the reference model, recognizing that the control effectiveness is highly dependent on the matching accuracy between the reference model and the closed--loop system.
It combines the advantages of both methods: ease of design by FRIT and the practical convenience of MPC.
This overcomes the limitations associated with FRIT's reliance on reference models and enhances the usability of MPC without significantly increasing the overall design effort.

In our previous study, the proposed method was applied experimentally \cite{IFAC}.
This study extends the foundational work by focusing on the integration with MPC for practical applications.
Specifically, the proposed method examines whether it can maintain the intuitive design aspects of traditional control while reducing design effort when combined with MPC.
Numerical simulations were conducted using two nonlinear model classes that exemplified structures commonly observed in industrial control systems.
These models feature both linear and nonlinear components that can effectively describe the hysteresis characteristics and dead zones.
The simulations not only demonstrate the applicability of the proposed method in practical settings but also provide technical insights through frequency--domain analysis.
Additionally, control experiments for angle tracking using artificial muscles with nonlinear characteristics were conducted using two types of reference signals.

The contributions of this study are as follows.

\begin{enumerate}
        \item Proposal and design of a PL method through systematic optimization combining FRIT and MPC for SISO systems;
        \item Evaluation of the control performance of the Bouc--Wen model and Hammerstein model via simulations for hysteresis and dead zone classes;
        \item Analysis of the proposed method from an MPC perspective;
        \item Experimental evaluation of the PL model and weight $R$ for artificial muscles with sinusoidal and square references.
\end{enumerate}

The remainder of this paper is organized as follows:
First, FRIT and its extended versions are briefly reviewed in Section 2.
Next, the PL method using FRIT is explained and the PL model is introduced in Section 3.1.
Subsequently, a design method for model--based control systems using the PL model is proposed.
The subsequent sections detail the design of the MPC, which explicitly considers control input constraints using the PL model in Section 3.2 and 3.3.
Finally, to confirm the effectiveness of the proposed method, reference tracking control was performed and analyzed through simulations for two nonlinear classes in Section 4 and experiments for artificial muscle actuators in Section 5. Furthermore, the obtained results were compared with those achieved using conventional methods.
    
\section{Data--driven Control}\label{sec2}

In this section, first, an overview of the conventional FRIT method and its principles is provided.
Next, the E--FRIT method, which improving control performance by suppressing the control input signal, is explained.

FRIT is a data--driven control method that determines controller parameters using only a single closed--loop I/O data.
\begin{figure}
\centering 
        \includegraphics[width=3.2in]{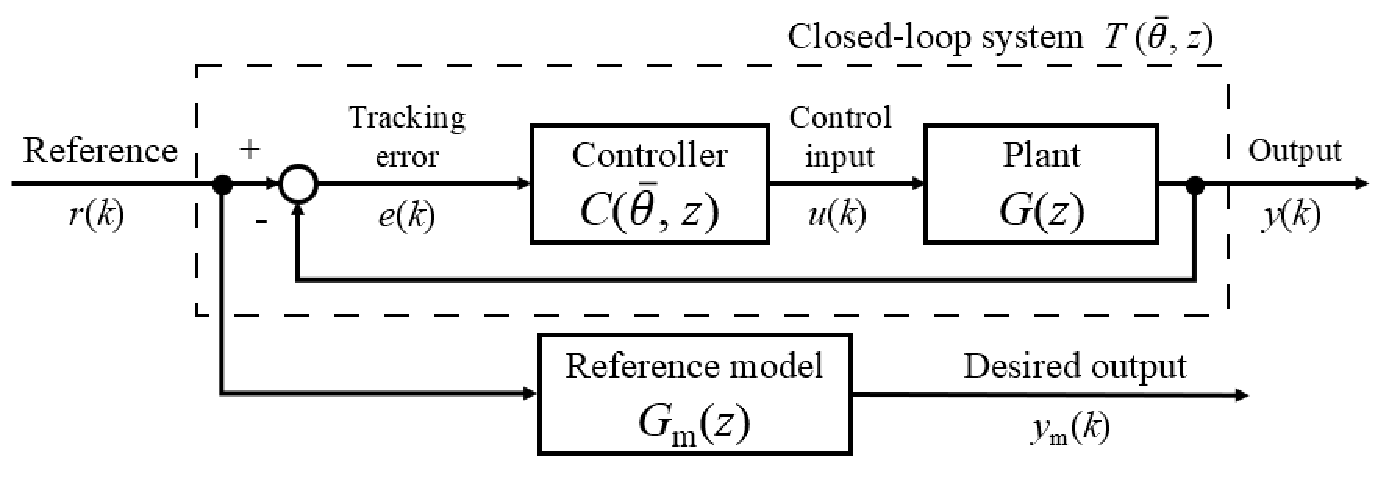}
        \caption{Block diagram for FRIT method.}
        \label{fig:closed_loop}
        \end{figure}
We introduce the time--shift operator $z$ and consider the block diagram shown in Fig. \ref{fig:closed_loop}, where $G(z)$ denotes an unknown plant system and $C(\bar{\bm{\theta}}, z)$ denotes a discrete--time PID controller structure expressed as follows:
\begin{equation}
        \label{Controller structure}
        C(\bar{\bm{\theta}}, z) = \bm{\beta}^{T}(z)\bar{\bm{\theta}} = K_{\rm P} + \frac{K_{\rm I}T_{\rm s}}{1 - z^{-1}} + \frac{K_{\rm D}(1 - z^{-1})}{T_{\rm s}},
\end{equation}
    where $\bm{\beta}(z) = [1, T_{\rm s}/(1-z^{-1}), (1-z^{-1})/T_{\rm s}]^{\rm T} \in \mathbb{R}^{3}$ denotes the known vector of the discrete--time transfer function, $\bar{\bm{\theta}} = [K_{\rm P}, K_{\rm I}, K_{\rm D}]^{\rm T} \in \mathbb{R}^{3}$ denotes the vector of the controller parameters, and $T_{\rm s}$ denotes the sampling time.
    Furthermore, $T(\bar{\bm{\theta}}, z)$ denotes a closed--loop system composed of $C(\bar{\bm{\theta}}, z)$ and $G(z)$.
    The objective of the FRIT method includes minimizing the following model reference evaluation function:
    \begin{equation}
        \label{evo}
        J(\bar{\bm{\theta}}) = \left\| \frac{G(z)C(\bar{\bm{\theta}},z)}{1 + G(z)C(\bar{\bm{\theta}},z)}r(k)-G_{\rm m}(z)r(k) \right\|^{2}_{2},
    \end{equation}
    where, $\|\cdot\|_{2}$ denotes the Euclidean norm, $G_{\rm m}(z)$ denotes the reference model appropriately selected by the designer, and $r(k) \in \mathbb{R}$ denotes the reference.
    As expressed in (\ref{evo}), FRIT is essentially a model--matching problem in which $\bar{\bm{\theta}}$ is numerically optimized to match $T(\bar{\bm{\theta}}, z)$ with $G_{\rm m}(z)$.
    To realize the aforementioned objective, the prior single I/O data: $y_{0}(\bar{\bm{\theta}}_{0}, k) \in \mathbb{R}$ and $u_{0}(\bar{\bm{\theta}}_{0}, k) \in \mathbb{R}$ in the closed--loop system obtained in a prior experiment with initial parameters $\bar{\bm{\theta}}_{0}$ are used to tune $\bar{\bm{\theta}}$.
    However, minimizing the evaluation function in (\ref{evo}) is impossible because $G(z)$ in (\ref{evo}) is an unknown system. Therefore, we rewrite the evaluation function (\ref{evo}) to (\ref{FRIT_evo1}) as follows:
    \begin{equation}
        \label{FRIT_evo1}
        J_{\rm F}(\bar{\bm{\theta}}) = \left\| y_{0}(\bar{\bm{\theta}}_{0}, k)-\tilde{y}(\bar{\bm{\theta}},k) \right\|^{2}_{2}.
    \end{equation}
    Here, $\tilde{y}(\bar{\bm{\theta}},k)$ denotes the desired output when $\tilde{r}(\bar{\bm{\theta}},k)$ is the input:
    \begin{equation}
        \label{FRIT_evo2}
        \tilde{y}(\bar{\bm{\theta}},k) = G_{\rm m}(z) \tilde{r}(\bar{\bm{\theta}},k),
    \end{equation}
where, $\tilde{r}(\bar{\bm{\theta}},k)$ denotes the fictitious reference signal:
    \begin{equation}
        \label{r_tilde}
        \tilde{r}(\bar{\bm{\theta}},k)=C^{-1}(\bar{\bm{\theta}},z)u_{0}(\bar{\bm{\theta}}_{0},k)+y_{0}(\bar{\bm{\theta}}_{0},k).
    \end{equation}
Here, $\tilde{r}(\bar{\bm{\theta}},k)$ can be calculated, because the evaluation function (\ref{FRIT_evo1}) does not contain an unknown system $G(z)$.
Therefore, the FRIT provides the desired controller parameters without requiring a mathematical model of the plant.
 Similarly, substituting $\tilde{r}(\bar{\bm{\theta}},k)$ into the closed--loop system $T({\bar{\bm{\theta}},z})$ with the parameter $\bar{\bm{\theta}}$, the initial data $y_{0}(\bar{\bm{\theta}}_{0},k)$ is reproduced as follows:
   \begin{equation}
       \label{y_ini_re}
      T({\bar{\bm{\theta}},z})\tilde{r}(\bar{\bm{\theta}},k) = \frac{G(z)C(\bar{\bm{\theta}},z)}{1 + G(z)C(\bar{\bm{\theta}},z)}\tilde{r}(\bar{\bm{\theta}},k) =y_{0}(\bar{\bm{\theta}}_{0},k),
    \end{equation}
    where, $y_{0}(\bar{\bm{\theta}}_{0},k) = G(z)u_{0}(\bar{\bm{\theta}}_{0},k)$.
 Furthermore, $\tilde{r}(\bar{\bm{\theta}},k) = \frac{1}{T({\bar{\bm{\theta}},z})}y_{0}(\bar{\bm{\theta}}_{0},k)$ according to (\ref{y_ini_re}). Substituting this into (\ref{FRIT_evo2}) and then incorporating the result into (\ref{FRIT_evo1}) yields (\ref{FRIT_evo_fc}).
    \begin{equation}
        \label{FRIT_evo_fc}
        J_{\rm F}(\bar{\bm{\theta}}) = \left\| \left( 1 - \frac{G_{\rm m}(z)}{T({\bar{\bm{\theta}},z})} \right)y_{0}(\bar{\bm{\theta}}_{0},k) \right\|^{2}_{2}
    \end{equation}
(\ref{FRIT_evo_fc}) can be transformed into the frequency domain using Parseval's theorem as follows:
    \begin{equation}
        \label{FRIT_evo_fc2}
        J_{\rm F}(\bar{\bm{\theta}}) \approx
        \frac{1}{2\pi}    \int_{- \pi}^{\pi}   \left\| \left( 1 - \frac{G_{\rm m}(e^{j\omega})}{T({\bar{\bm{\theta}},e^{j\omega}})} \right) Y_{0}(\bar{\bm{\theta}}_{0},e^{j\omega}) \right\|^{2}_{2} d\omega,
    \end{equation}
 where, $Y_{0}(\bar{\bm{\theta}}_{0}, e^{j\omega})$ denotes the power spectral density of $y_{0}(\bar{\bm{\theta}}_{0}, k)$.
It can be observed that the evaluation function of FRIT numerically determines the value of $\bar{\bm{\theta}}$ such that $T(\bar{\bm{\theta}}, z)$ matches $G_{\rm m}(z)$ for the frequency components contained in the initial output $y_{0}(\bar{\bm{\theta}}_{0}, k)$.
Because $Y_{0}(\bar{\bm{\theta}}_{0}, e^{j\omega})$ depends on the reference $r_{0}(k)$ in the preliminary experiment and the initial parameters $\bar{\bm{\theta}}_{0}$, the FRIT method performs local matching of $G_{\rm m}(z)$ and $T(\bar{\bm{\theta}}, z)$ under initial conditions that depend on these two factors.
Next, the evaluation function for E--FRIT, which effectively suppresses the input variations, is detailed.
The E--FRIT evaluation function is as follows\cite{E-FRIT}:
    \begin{equation}
        \label{EFRIT_EVO}
        J_{\rm EF}(\bar{\bm{\theta}}) = J_{\rm F}(\bar{\bm{\theta}}) + \lambda  \left\| \Delta \tilde{u}(\bar{\bm{\theta}}, k)\right\|^{2}_{2},
    \end{equation}
    where, $\lambda$ denotes indicates the weight and $\Delta \tilde{u}(\bar{\bm{\theta}}, k)$ denotes the fictitious input variation.
    \begin{eqnarray}
        \label{Delta_utilde}
        \Delta \tilde{u}(\bar{\bm{\theta}},k)  &\triangleq&  \tilde{u}(\bar{\bm{\theta}},k) - \tilde{u}(\bar{\bm{\theta}},k-1) \\
        \label{utilde}
        \tilde{u}(\bar{\bm{\theta}},k)  &=& C(\bar{\bm{\theta}},z)( \tilde{r}(\bar{\bm{\theta}},k) -  \tilde{y}(\bar{\bm{\theta}},k))
    \end{eqnarray}
\begin{remark}
This method does not readily adapt to changes in system characteristics or variations in the reference frequency during operation.
Moreover, it fails to reflect the designer's preferences and the input constraints of the system as effectively as those reflected by MBC methods and adaptive control strategies.
\end{remark}
\begin{remark}
The performance of the designed controllers using the E--FRIT method depends significantly on the choice of the weighting parameter $\lambda$ in (\ref{EFRIT_EVO}).
An appropriately chosen $\lambda$ can lead to performance improvement by effectively managing the tradeoff between control accuracy and robustness.
However, determining the optimal $\lambda$ is nearly as complex as tuning the reference model.
\end{remark}
    
\section{Model-based Control System Using PL}
\begin{figure*}[!t]
        \centering
        \includegraphics[width=5in]{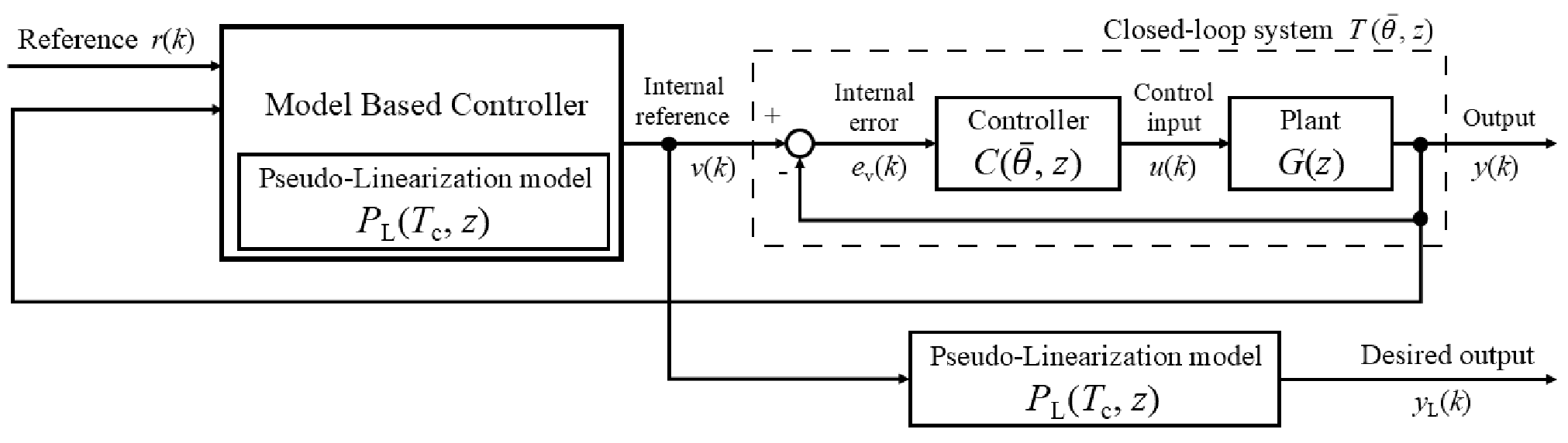}
        \caption{Block diagram for model--based controller using PL with E--FRIT}
        \label{fig:model_based_pl}
    \end{figure*}

  In this section, we propose a method for designing model--based control systems using the PL technique based on E--FRIT.
    This controller method is applicable to linear and nonlinear time--invariant SISO systems.
    A block diagram of the proposed method is presented in Fig. \ref{fig:model_based_pl}, where $v(k) \in \mathbb{R}$ denotes the input generated by the model--based controller and $u(k)$ denotes the actual input for plant via PID controller $C(\bar{\bm{\theta}}, z)$.
    Furthermore, $e_{\rm v}(k) \in \mathbb{R}$ denotes the internal error between $v(k)$ and $y(k)$,
    %
    %
    %
    %
    Typically, the purpose of E--FRIT is to perform model matching between an augmented system and a reference model corresponding to the ideal model for a given reference signal.
    However, the optimal design of an appropriate reference model for unknown plants remains challenging.
    In Section 3.1, we address this issue by explicitly focusing on E--FRIT.
    We propose a method for separating E--FRIT from the reference tracking problem and introduce a PL technique that allows a closed-- loop system to match a linear model with adjustable parameters.
    The linear model used for model matching was defined as the PL model.
    The PL model differs from the reference model because it does not use ideal characteristics as references.
    Furthermore, because the PL model was optimized to match the closed--loop system, it was not designed to consider its relationship with the reference.

    Subsequently, a model--based controller was designed in the outer loop using the obtained PL model, as shown in Fig. \ref{fig:model_based_pl}.
    However, the PL model depends solely on the I/O data and cannot explicitly handle the control input $u(k)$ in $G(z)$, which matches the state variables of the closed--loop system.
    To address this issue, a method for estimating the control input $u(k)$ is proposed in Section 3.2.
    It uses a PL model and a PID controller based on the internal error $e_{\rm v}(k)$.
    Finally, in Section 3.3, a MPC system is designed using the PL model as a predictor, and the input $u(k)$ is estimated by considering the input constraints.
\subsection{PL using E--FRIT}
    In this section, we propose a PL technique for unknown systems using the E--FRIT method.
    As discussed earlier, E--FRIT designs a controller that matches the closed--loop system $T(\bar{\bm{\theta}}, z)$ with a reference model $G_{\rm m}(z)$ that ideally exhibits the desired system characteristics.
    However, in the absence of information on $G(z)$, designing a reference model appropriately is challenging.
    Moreover, forcing an unknown $T(\bar{\bm{\theta}}, z)$ to match an inappropriate $G_{\rm m}(z)$ can degrade system control performance.
    %
    %
    Therefore, to improve control performance, it is necessary to optimize the reference model.
    However, this approach involves arbitrary alteration of the desired output.
    Consequently, it cannot be guaranteed that the control system designed according to the reference model optimized using this approach will result in an improved control performance with respect to the reference.
    To address this challenge, we shift our focus from the fundamental concept of E--FRIT to model matching, thereby distinguishing it from reference tracking, and thus, the closed--loop system matches the PL model.
    In the proposed method, the E--FRIT method tunes $\bar{\bm{\theta}}$ such that the I/O characteristics of $T(\bar{\bm{\theta}},z)$ match the PL model. It does not aim to match the state variables of $T(\bar{\bm{\theta}},z)$ to those of the PL model.
    Therefore, in the control system designed using the PL model, the state variables fed back from $T(\bar{\bm{\theta}},z)$ do not have any physical interpretation other than the output signal.
    Hence, it is preferable for the PL model to have only one state variable.
    Therefore, we adopt a structure that is a discretization of the first--order system in (\ref{pl_model}):
    \begin{equation}
        \label{pl_model}
        P_{\rm L}(T_{\rm c},z) = \frac{ \left( 1-e^{-\frac{T_{\rm s}}{T_{\rm c}}} \right) z^{-1}}{1-e^{-\frac{T_{\rm s}}{T_{\rm c}}}z^{-1}},
    \end{equation}
    where, $T_{\rm c}$ is the time constant. Therefore, the adjustable parameter $\bar{\bm{\theta}}$ of the conventional E--FRIT using a PID controller including the adjustable parameter $T_{\rm c}$ of the PL model, can be rewritten as follows:
    \begin{equation}
        \label{PL_theta}
        \bm{\theta} = [ K_{\rm P}, K_{\rm I}, K_{\rm D}, T_{\rm c} ]^{\rm T} = [\bar{\bm{\theta}}^{\rm T}, T_{\rm c}]^{\rm T},
    \end{equation}
    where, $\bm{\theta}$ denotes the modified E--FRIT parameter used in the proposed method.
    This allowed us to obtain an optimized PL model for the plant and tune the control parameters.
    From (\ref{EFRIT_EVO}) and (\ref{PL_theta}), it can be observed that the closed--loop system composed of the obtained PL model and controller is separated from the reference tracking problem. This is because the PL model is used only for matching, and thus, it can be arbitrarily tuned through optimization.
    %
    %
    %
\begin{remark}
    When designing a PL model of a higher order than a first--order system, the model construction approach can vary significantly based on the availability of measurements for all state variables.
    In cases where all state variables are measurable, it is possible to employ a nonminimal state--space representation \cite{NMSS} to construct the PL model.
    However, for cases in which not all state variables can be directly measured, estimating the unobserved states becomes necessary.
    Techniques, such as observers or Kalman filters, are typically used for state estimation in these situations.
    However, as mentioned previously, states other than the output of the PL model do not have any physical interpretations.
    Thus, the information obtained from their estimation lacks significance, and the design of a PL model using observers remains an open problem.
\end{remark}
\subsection{Estimation of Internal Variables}
    The PL model is obtained solely from single I/O data without considering the system characteristics.
    This allows us to design various model--based controllers easily without time--consuming modeling routines.
    However, the state variables of the PL model lack inherent meaning, and it is impossible to directly consider the input $u(k)$ in $G(z)$, which corresponds to the state variables in the matched closed--loop system $T(\bar{\bm{\theta}}, z)$.
    Some model--based control approaches, such as MPC, generate control inputs by considering input constraints.
    However, in the proposed method, the controller generates input $v(k)$ as a reference for the inner--loop system, as shown in Fig. \ref{fig:model_based_pl}.
    This differs from the actual input $u(k)$ applied to a plant.
    To address this problem, we estimate the input $u(k)$ applied to the plant using the internal reference $v(k)$ generated by the model--based controller via the PL model and the tuned PID controller.
    The following state--space model is considered corresponding to the discretized PL model of (\ref{pl_model}):
    \begin{equation}
        \label{PL_state_model}
        \left\{
        \begin{aligned}
            x(k+1) &= a_{\rm P}x(k)+b_{\rm P}v(k) \\
            y(k) &= c_{\rm P}x(k)
        \end{aligned}
        \right.
        ,
    \end{equation}
    where, $x(k)$, $v(k)$, and $y(k) \in \mathbb{R}$ denote the state variable, internal reference used as input for the PL model, and the output, respectively.
    The parameters of the state--space representation are as follows:
    \begin{equation}
        \label{PL_state_model_parame}
        a_{\rm P} = e^{- \frac{T_{\rm s}}{T_{\rm c}}},\quad b_{\rm P} = 1- e^{- \frac{T_{\rm s}}{T_{\rm c}} }, \quad c_{\rm P} = 1
        .
    \end{equation}
    The state vector of $T(\bar{\bm{\theta}}, z)$, which is linearized to the PL model using E--FRIT, only corresponds to the plant output, given that E--FRIT tunes $\bm{\theta}$ such that $T(\bar{\bm{\theta}}, z)$ and the I/O characteristics of the PL model match.
    Therefore, it is preferable to include only the scalar state variable for the PL model handled by MPC.
    Hence, we assume that the predictor of the MPC is a first--order system, which is the PL model structure set by the designer.
    In this study, we estimate the $i$--step ahead input as $\hat{u}(k+i)$ from the PID controller and optimized PL model as follows:
    \begin{equation}
        \label{estimate_input}
        \hat{u}(k+i) = \hat{u}_{\rm P}(k+i) + \hat{u}_{\rm I}(k+i) + \hat{u}_{\rm D}(k+i),
    \end{equation}
    where, $\hat{u}_{\rm P}(k+i)$, $\hat{u}_{\rm I}(k+i)$, and $\hat{u}_{\rm D}(k+i)$ denote the estimated proportional, integral, and differential inputs, respectively, expressed as follows:
    \begin{equation}
        \label{estimate_input_PID}
        \left\{
        \begin{aligned}
            \hat{u}_{\rm P}(k+i)&= K_{\rm P} \hat{e}_{\rm v}(k+i)\\
            \hat{u}_{\rm I}(k+i)&=\hat{u}_{\rm I}(k+i-1)+K_{\rm I}\hat{e}_{\rm v}(k+i) T_{\rm s}\\
            \hat{u}_{\rm D}(k+i)&= K_{\rm D} \frac{\hat{e}_{\rm v}(k+i)-\hat{e}_{\rm v}(k+i-1)}{T_{\rm s}} \\
        \end{aligned}
        \right.
        ,
    \end{equation}
    Here, $\hat{e}_{\rm v}(k+i)$ denotes the estimation of the internal error using the estimated output $\hat{y}(k+i)$ of the PL model, which is expressed as follows:
    \begin{equation}
        \label{estimate_internal_error}
        \hat{e}_{\rm v}(k+i) = v(k+i) - \hat{y}(k+i),
    \end{equation}
    where, $v(k)$ is the internal reference generated by the model--based controller.
    Note that $\hat{u}(k)$ at $i=0$ because the real output can be available from (\ref{estimate_input})--(\ref{estimate_internal_error}).
    Furthermore, the estimation accuracy of $\hat{u}(k+i)$ for $ i \geq 1$ depends on the matching accuracy between the PL model and $T(\bar{\bm{\theta}}, z)$.
\subsection{MPC Design with PL Model}
    In this subsection, we design an MPC system as an example of a model--based control system using the PL model and the estimated input to the plant explicitly.
    Figure \ref{fig:MPC_based_pl} shows a block diagram of the proposed method, which uses the PL model as a predictor of the MPC.
    Hence, MPC can generate the optimal input while predicting the output of the closed--loop system.
    The cost function of the outer loop MPC shown in Fig. \ref{fig:MPC_based_pl} is expressed as follows:
    %
    %
    \begin{figure}
        \centering
        \includegraphics[width=3.5in]{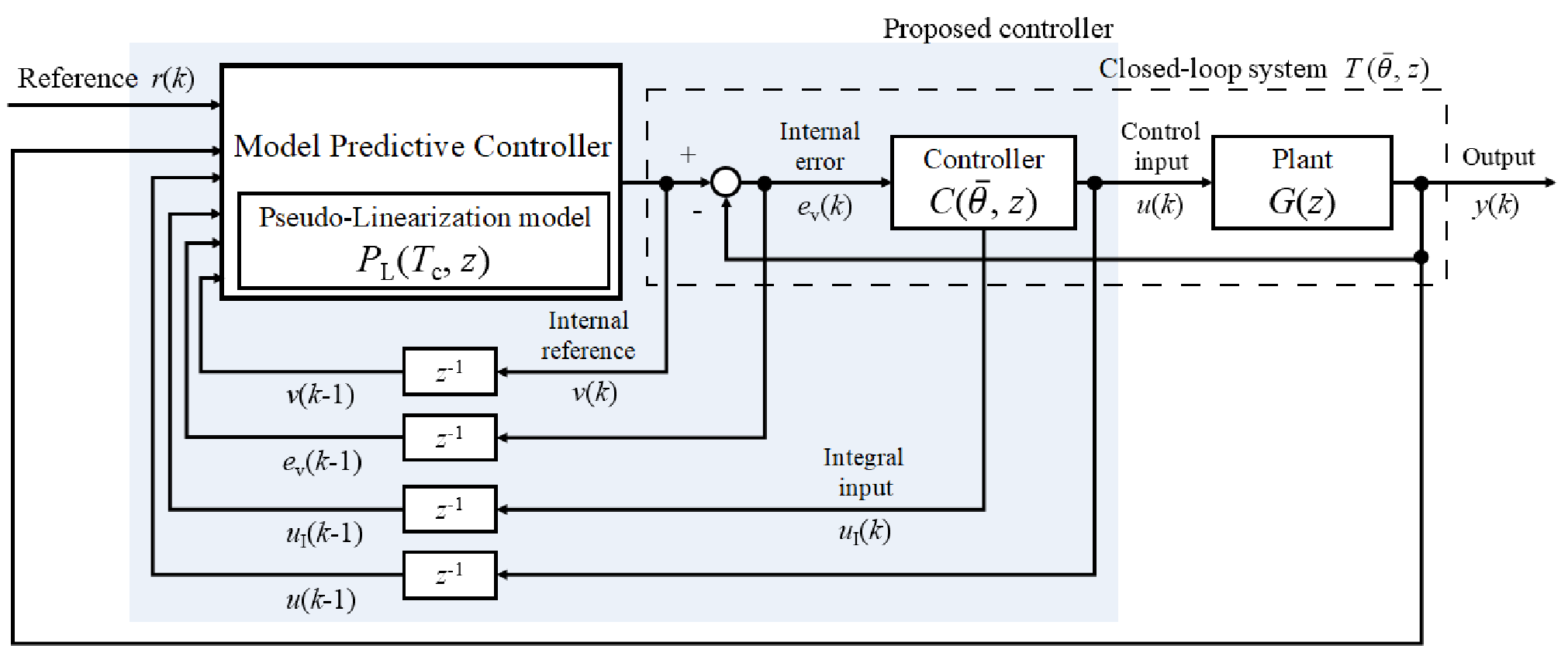}
        \caption{Block diagram for adaptive E--FRIT--based MPC}
        \label{fig:MPC_based_pl}
    \end{figure}
    %
    %
        \begin{align} 
        \label{MPC_evo}
            J(k,\Delta v) &= J_{y}(k,\Delta v) + J_{u}(k,\Delta v) +  J_{v}(k,\Delta v)\\
            &\hspace{10pt}\mathrm{subject \hspace{5pt} to} \hspace{5pt} u_{\rm min} \leq \hat{u}(k+i) \leq u_{\rm max}, \hspace{5pt} \notag\\
            &\hspace{50pt}\forall k \geq  0 ,\ i=0,\ldots, H_{\rm u}-1,\notag      
        \end{align} 
    and
    \begin{equation}
        \label{MPC_evo_term}
        \left\{
        \begin{aligned}
            J_{y}(k,\Delta v)&=\sum_{i=1}^{H_{\rm p}}\|\hat{y}(k+i|k)-r(k+i|k)\|^{2}_{Q(i)}\\
            J_{u}(k,\Delta v)&=\sum_{i=0}^{H_{\rm{u}}-1} \|\Delta\hat{u}(k+i|k)\|^{2}_{R(i)}\\
            J_{v}(k,\Delta v)&=\sum_{i=0}^{H_{\rm{u}}-1} \|\Delta{v}(k+i|k)\|^{2}_{V(i)}
        \end{aligned}
        \right.
        .
    \end{equation}
    where, $J_{y}(k,\Delta v)$, $J_{u}(k,\Delta v)$, and $J_{v}(k,\Delta v)$ denote the evaluation terms for the error, input variation, and internal reference variation, respectively.
    $H_{\rm p}$ and $H_{\rm u}$ denote the prediction and control horizons, respectively.
    $J_{u}(k,\Delta v)$ evaluates $\Delta v(k)$ as a variable of the cost function, because $\Delta \hat{u}(k)$ is generated by $\Delta v(k)$ as indicated by (\ref{estimate_input})--(\ref{estimate_internal_error}).
    Furthermore, $u_{\rm min}$ and $u_{\rm max}$ denote the input constraints of the plant, and the matrices $Q(i) \geq 0$ $(i=1,\ldots,H_{\rm p})$, $R(j) \geq 0$, $V(j) > 0$ $(j=0,\ldots,H_{\rm
        u}-1)$ denote the weight of each evaluation term.
    In the proposed method, the optimized variable that the MPC handles is $v(k)$, and the internal reference variation weight $V(i)$ is provided as a positive definite matrix.
    Therefore, similar to the general MPC weight design, $Q(i)$ is tuned to reduce the error and $V(i)$ is tuned to attenuate excessive input variation.
    In addition, the input variation weight $R(i)$ is tuned to suppress the oscillation of the input $u(k)$ applied to the plant.
    In cases where oscillations did not occur, $R(i)$ is designed to be zero.
    Specifically, in the simulations described in Section 4, $R(i)\equiv0$, $\forall i$ were set.
    For simplicity, it is assumed that $H_{\rm p}=H_{\rm u}$.
    Therefore, the result of the MPC under constraints provides an optimized internal reference $v(k)$ by minimizing the evaluation function in (\ref{MPC_evo}), allowing high control performance to be achieved by considering the plant input constraints without plant model information.
    Here, we present an algorithm that outlines the design of the proposed method.
   It provides a structured approach for coordinating the principles of PL and MPC, as detailed in Algorithm \ref{alg:design_methodology}.

    In the following sections, we validate the proposed method thoroughly using simulations and experimental evaluations.
    Specifically, in Section 4, simulations are conducted to illustrate the efficacy of our method in handling two distinct classes of nonlinearity.
    Through these simulations, we aim to clarify the performance of the proposed method in scenarios characterized by hysteresis and dead zones.
    In Section 5, the experimental results of tap--water--driven artificial muscle control systems are examined.

\begin{algorithm}
\caption{Design Methodology}
\label{alg:design_methodology}
\begin{algorithmic}[]
    \State \textbf{Offline Preprocessing:}
    \State \textbf{Step 1:} Obtain the I/O data\\$u_0(\bar{\bm{\theta}}_{0},k)$, $y_0(\bar{\bm{\theta}}_{0},k)$, ($k = 1, \cdots, N$) of plant in closed-loop data.
    \State \textbf{Step 2:} Calculate:
    \State $\cdot$ Obtain a fictitious reference signal $\tilde{r}(\bar{\bm{\theta}},k)$ as follows:
    \State \hspace{10mm}$\tilde{r}(\bar{\bm{\theta}},k)=C^{-1}(\bar{\bm{\theta}},z)u_{0}(\bar{\bm{\theta}}_{0},k)+y_{0}(\bar{\bm{\theta}}_{0},k)$
    \State $\cdot$ Calculate the desired output $\tilde{y}(\bm{\theta},k)$ as follows:
    \State \hspace{10mm}$\tilde{y}(\bm{\theta},k) = P_{\rm L}(T_{\rm c}, z) \tilde{r}(\bar{\bm{\theta}},k)$
    \State $\cdot$ Calculate the fictitious input $\tilde{u}(\bm{\theta},k)$ as follows:
    \State \hspace{10mm}$\tilde{u}(\bm{\theta},k) = C(\bar{\bm{\theta}},z)( \tilde{r}(\bar{\bm{\theta}},k) -  \tilde{y}(\bm{\theta},k))$
    \State $\cdot$ Obtain the fictitious input variation $\Delta \tilde{u}(\bm{\theta},k)$ as follows:
    \State \hspace{10mm}$ \Delta \tilde{u}(\bm{\theta},k) \triangleq \tilde{u}(\bm{\theta},k) - \tilde{u}(\bm{\theta},k-1)$
    \State \textbf{Step 3:} Set a weight $\lambda$
    \State \textbf{Step 4:} Minimize the following evaluation function:
    \State \hspace{10mm}$ J_{\rm EF}(\bm{\theta}) = J_{\rm F}(\bm{\theta}) + \lambda  \left\| \Delta \tilde{u}(\bm{\theta}, k)\right\|^{2}_{2}$
    \State \hspace{10mm}where, $J_{\rm F}(\bm{\theta}) = \left\| y_{0}(\bar{\bm{\theta}}_{0}, k)-\tilde{y}(\bm{\theta},k) \right\|^{2}_{2}$
    \State \textbf{Step 5:} Design the linear MPC using the optimized PL model as the predictor, considering input constraints\\
    \State \textbf{Online Processing:}
    \State \textbf{Step 6:} Estimate control input $\hat{u}(k+i)$\\ and internal states using the designed PL model by (\ref{estimate_input})--(\ref{estimate_internal_error})
    \State \textbf{Step 7:} Minimize the evaluation function $J(k,\Delta v)$ in (\ref{MPC_evo}) to obtain the optimal control input $v(k)$
    \State \textbf{Step 8:} Set $k \leftarrow k+1$ and go to \textbf{Step 6}
\end{algorithmic}
\end{algorithm}
 \section{Simulation Study}
    In this section, the effectiveness of the proposed method is verified using two important classes of nonlinearity:
    Hammerstein model and Bouc--Wen model.
    The Hammerstein model is frequently used as a case study for data--driven control \cite{hammer1,hammer2} to confirm the effectiveness of the proposed method for nonlinear classes.
    In contrast, the asymmetric Bouc--Wen model denotes the characteristics of the artificial muscles discussed in Section 5.
    Therefore, the simulations correspond to the experiment presented in Section 5 and are suitable for confirming the effectiveness of the proposed method and for analyzing the frequency domain, which cannot be confirmed experimentally.
    In addition, we conducted simulations that focused on frequency domain matching for PL using the proposed E--FRIT method.
    For the asymmetric Bouc--Wen model, which can denote the hysteresis characteristics, we verified the matching between the frequency characteristics of the designed PL model and those of the designed closed--loop system.
    \subsection{Simulation: Hammerstein Model}
    To verify the effectiveness and advantages of the proposed method over the conventional E--FRIT method, a simulation was conducted using the following second--order Hammerstein model:
    The system model presented in \cite{hammer1,hammer2} is expressed as follows:
    \begin{equation}
        \label{Hammer}
        \left\{
        \begin{aligned}
            y(k) =\ & 0.6y(k-1) - 0.1y(k-2)  \\
            &+1.2x(k-1) - 0.1x(k-2) \\
            x(k) =\ &1.5u(k)-1.5u^{2}(k) + 0.5u^{3}(k)\\
        \end{aligned}
        \right.
        .
    \end{equation}
    Reference $r(k)$ is given by
    \begin{equation}
        \label{Hammer_ref}
        r(k)=
        \left\{
        \begin{aligned}
            0.5, \quad &0 \leq k < 50\\
            1.0, \quad  &50 \leq k < 100\\
            2.0, \quad  &100 \leq k < 150\\
            1.5, \quad  &150 \leq k < 200\\
        \end{aligned}
        \right.
        .
    \end{equation}
    The model in (\ref{Hammer}) was used solely to generate I/O data.
    The controller configuration used in the proposed method is shown in Fig. \ref{fig:MPC_based_pl}, while the conventional method uses the controller configuration shown in Fig. \ref{fig:closed_loop}. The initial PID gains were set to $\bar{\bm{\theta}}_{0} = [1.00\times 10^{-2},1.00\times 10^{-2},1.00\times 10^{-3}]^{\rm T}$, and the controller was designed using the parameter set $\bm{\theta}$ obtained through the PL with E--FRIT.
    Using $\lambda = 1.00\times10^{3}$, the following parameters were obtained through optimization: $\bm{\theta} = [4.71\times10^{-9},9.09\times10^{-1},3.68\times10^{-11},0.81]^{\rm T}$.
    In the conventional method, the controller was designed using a PL model that incorporated $T_{\rm c}$.
    In contrast, in the proposed method, the parameters of the PL model were adopted as predictions for the MPC.
    Both the prediction horizon $H_{\rm p}$ and the control horizon $H_{\rm u}$ were set to 5 steps.
    In this simulation, we used two distinct weight settings for an efficient analysis of the proposed method from the MPC perspective.
    This approach aims to evaluate the capability of the method to reflect the designer's preferences in the system design, which is a result of integrating E--FRIT with MPC.
    The weight setting of Case 1 emphasizes the rapid tracking performance, with weight matrices defined as $Q(i)\equiv 1000I_{5}$, $R(i)\equiv 0$, and $V(i)\equiv I_{5}$, $\forall i$.
    However, in the weight setting of Case 2, the focus shifts to the steady--state tracking performance, setting the weight matrices to $Q(i)\equiv I_{5}$, $R(i)\equiv 0$ and $V(i)\equiv 100I_{5}$, $\forall i$.
    This configuration explores the extent to which designers can reflect their preferences under specific operational conditions.
    As mentioned in Section 3.3, the weight matrices $R(i)$ were set to zero in the simulation.
    Furthermore, the input constraints were set to $u_{\rm min}=0$ and $u_{\rm max}=2$ to prevent excessive inputs in the proposed method.
    Real--time control code was generated using the common convex optimization solver CVXGEN \cite{CVXGEN}.
    %
    %

\begin{figure}
        \centering
        \includegraphics[width=3in]{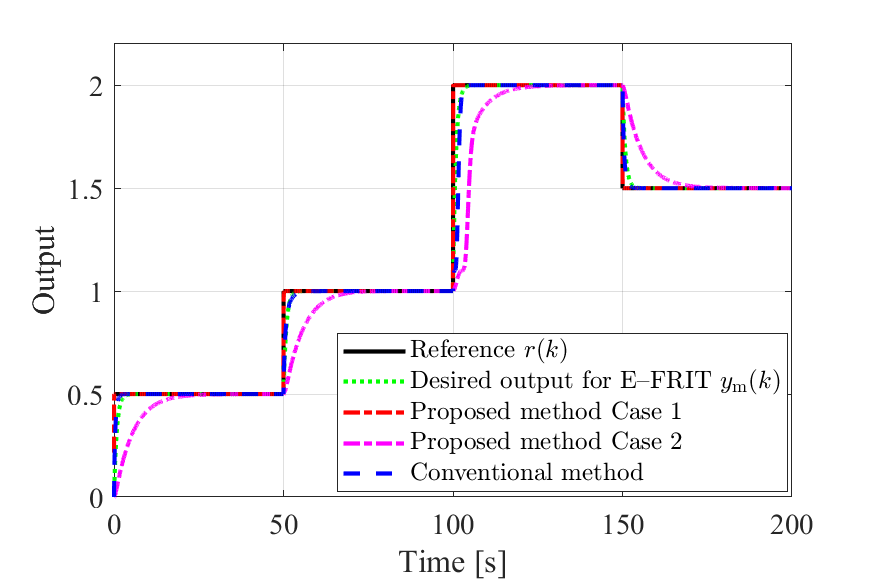}
        \caption{Comparison of tracking performance between the proposed and conventional methods in simulations of the Hammerstein model}
        \label{fig:hammer_result}
    \end{figure}

\begin{figure}
        \centering
        \includegraphics[width=3in]{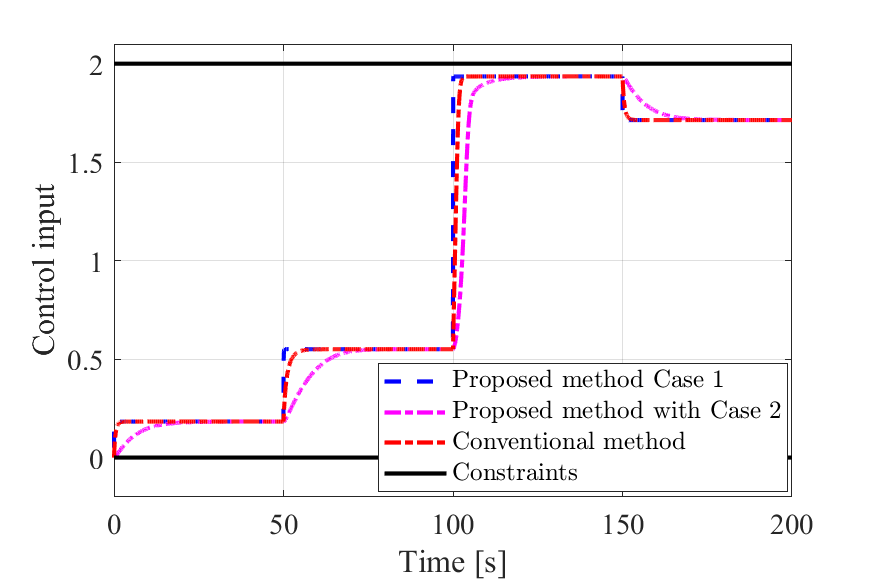}
        \caption{Control input $u(k)$ to the plant in simulations of the Hammerstein model}
        \label{fig:hammer_input}
    \end{figure}

    The simulation results are shown in Figs. \ref{fig:hammer_result} and \ref{fig:hammer_input}.
    As shown in Fig. \ref{fig:hammer_result}, the conventional method gradually tracked the reference.
    This is because the desired output $y_{\rm m}(k)$ was designed according to $P_{\rm L}(z)$ which was optimized without considering a reference, and the control system was designed to track $y_{\rm m}(k)$.
    On the other hand, the proposed method tracked the reference $r(k)$ accurately.
    As the control performance of the conventional method strongly depends on the design of $T_{\rm c}$, it is generally difficult to design a $P_{\rm L}(z)$ that exhibits the same performance as the proposed method.
    In particular, in Case 1, which emphasizes the rapid tracking performance with weight matrices, the proposed method achieved a quick response by considering the input constraints effectively to avoid the wind--up phenomenon, as shown in Fig. \ref{fig:hammer_input}.
    Conversely, Case 2 focused on the long--term tracking performance, reflecting the designer's preference through a balanced setting of weight matrices.
    Therefore, in the proposed method, the control performance does not depend on $T_{\rm c}$ and a designer can intuitively design the control performance by tuning certain weights of the cost function, as in MPC.
    The average root mean square error (RMSE) of the output was $1.16 \times 10^{-2}$ for the proposed method in Case 1, showing its effectiveness compared with the RMSE $8.18 \times 10^{-2}$ for the conventional method.

     \subsection{Asymmetric Bouc--Wen Model}
    Several asymmetric Bouc--Wen models have been proposed to describe the hysteresis characteristics of plants \cite{BW,ABW}.
    In this study, we used the previously proposed asymmetric Bouc--Wen model \cite{muscle_MBD1} because of its simple structure and high accuracy.
    (\ref{Asymmetric Bouc--Wen}) expresses the asymmetric Bouc--Wen model.
    \begin{equation}
        \label{Asymmetric Bouc--Wen}
        \left\{
        \begin{aligned}
            y(k) =& \hspace{1mm} Y(k) + h(k) \\
            Y(k) =& \hspace{1mm} a_{1}y(k-1) + a_{2}y(k-2)+b_{2}u(k-1)\\
            h(k) =& \hspace{1mm} \sum_{i=1}^{2} h_{i}(k-i)\\
            h_{i}(k-i) =&\hspace{1mm} A_{i} \{y(k-i) - y(k-i-1)\} \\
            &+\beta_{i} |y(k-i)-y(k-i-1)|h_{i}(k-i) \\
            &+\gamma_{i}\{y(k-i) - y(k-i-1) \} |h_{i}(k-1)|\\
            &+c_{i}h(k-i)+d_{i}y^{2}(k-i)+e_{i}y^{3}(k-i),\\
            &&\!\!\!\!\!\!\!\!\!\!\!\!\!\!\!\!\!\!\!\!\!\!\!\!\!\!\!\!\!\!\!\!\!\!\!\!\!\!\!\!\!\!\!\!\!\!\!\!\!\! i = 1,2
        \end{aligned}
        \right.
        .
    \end{equation}
    where, $a_{1}$, $a_{2}$, and $b_{1}$ are linear parameters, and $A_{i}$, $\beta_{i}$, $\gamma_{i}$, $c_{i}$, $d_{i}$, and $e_{i}$ $(i \! = \! 1,2)$ are hysteresis parameters.
    The nonlinear function $h(k)$ denotes a virtual hysteresis variable.
    The asymmetric Bouc--Wen model can accurately denote the asymmetric hysteresis, such as that observed in artificial muscles \cite{muscle_MBD1}.
    Therefore, the proposed method can be analyzed using this model for a simulation study of the experimental setup described later.
    %
    %
    \subsection{Experimental Setup and Simulation Conditions}
\begin{figure}
        \centering
        \includegraphics[width=2.5in]{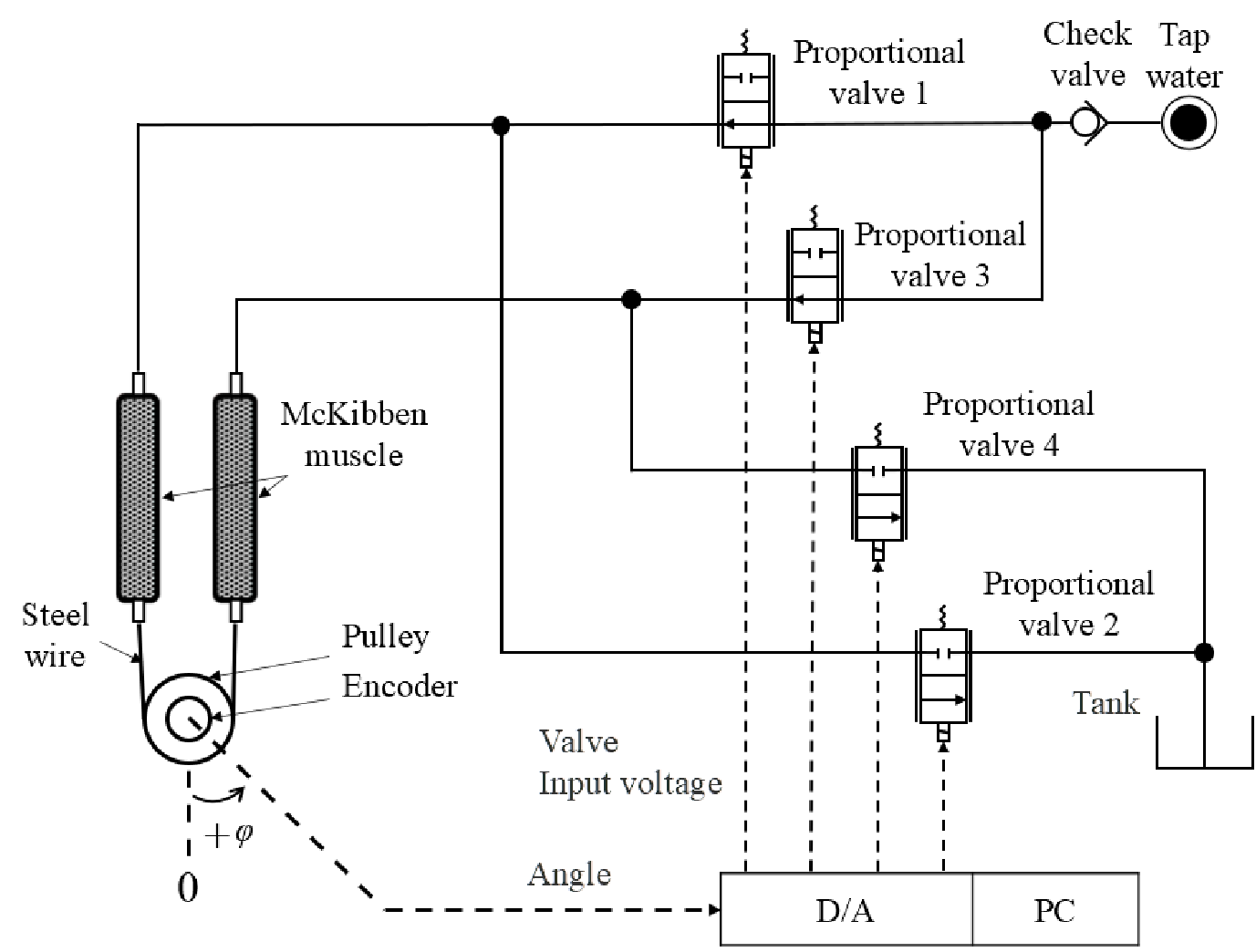}
        \caption{Experimental circuit.}
        \label{fig:Experimental circuit}
    \end{figure}

\begin{figure}
        \centering
        \includegraphics[width=2.4in]{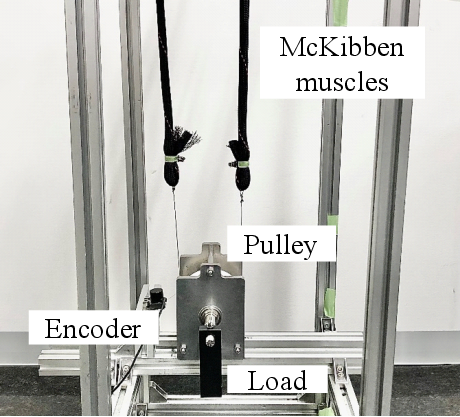}
        \caption{Experimental equipment.}
        \label{fig:Experimental equipment}
    \end{figure}

    We now discuss the parameters of the asymmetric Bouc--Wen model used in the simulations.
    To obtain the system parameters, system identification was performed using a tap--water--driven artificial muscle actuator, as shown in Figs. \ref{fig:Experimental circuit} and \ref{fig:Experimental equipment} and Table \ref{table:Experimental}.
    Artificial muscles have attracted considerable attention because of their advantages, such as low cost, light weight, high power density, high flexibility, and simple structure \cite{muscle}.
    In particular, tap--water--driven artificial muscles have advantages as aqua drive systems (ADSs), owing to the low environmental load, availability, and disposability of the pressure medium \cite{ADS}.
    At various pressure levels of an ADS, tap water is easily used because it does not require special equipment such as a compressor, power supply, or reservoir tank \cite{ADS_muscle}.
    However, the muscles have strong asymmetric hysteresis owing to the nonlinear contraction behavior and friction between the components.
    Therefore, controlling artificial muscles with high precision is challenging \cite{muscle_MBD1,muscle_MBD2}.
    Several methods using data--driven control have been applied to the muscles \cite{muscle_FRIT,muscle_MFAC}.
    The proposed method is superior to model--based controllers \cite{muscle_MBD1,muscle_MBD2} because it does not require explicit mathematical models.
    Furthermore, when compared to other data--driven controllers, it has the advantage of considering input constraints in the same manner as in MPC \cite{muscle_FRIT,muscle_MFAC}.

    The experimental setup consists of two McKibben--type artificial muscles, four proportional valves, a pulley, a wire, an encoder, and a controller PC.
    The muscles contract under tap water pressure.
    During the operation, one muscle is pressurized and the other is relaxed according to the valve opening, generating a pressure difference.
    This results in a difference in the tension between the two artificial muscles, which drives the wire and causes the pulley connected to the wire to rotate.
    Proportional valves are a type of flow control in which the flow rates are controlled by the input voltage $u(k)$.
    The encoder measures the pulley at a rotational angle of $y(k)$ and sends it to the controller.
    Because the system is configured to generate inputs for the four proportional valves using a single input, it can be considered a nonlinear SISO system.
    Details of the experimental setup are presented in Table \ref{table:Experimental}.
    \begin{table}
        \begin{center}
            \caption{Specifications of experimental components.}
            \label{table:Experimental}
            \begin{tabular}{ll}\hline
                \rule[0mm]{0mm}{4mm}Item & Specifications\rule[0mm]{0mm}{4mm}\\\hline\hline
                \rule[0mm]{0mm}{4mm}Proportional valves & KFPV300-2-80,\\
                & Koganei Corporation. \\
                & $C_v$ Value: 1.6;\\
                & Range of input voltage: 0 to 10 V\\
                & \\
                Linear encoder & DX-025, MUTOH Industries Ltd. \\
                & Resolution: 0.01 mm\\
                & \\
                Controller PC & Operating system: Windows 10,\\
                & Microsoft Corporation. \\
                & CPU: 2.50 GHz,\\
                & RAM: 16.00 GB\\
                & Applications: MATLAB/Simulink\\
                & and dSPACE 1103\\
                & \\
                Tap--water--driven & Handmade muscle,\\ muscle & Length: 400 mm\\
                & \\
                Tap--water & Average supply pressure:\\
                & 0.15 MPa (G)\\\hline
            \end{tabular}
        \end{center}
    \end{table}
    We conducted system identification under these specifications.
    The parameter sets obtained are listed in Table \ref{table:parameter set}.
    The method used for system identification is based on that described in \cite{muscle_MBD1}.
    \begin{table}
        \begin{center}
            \caption{Parameter set of the asymmetric Bouc--Wen model.}
            \label{table:parameter set}
            \begin{tabular}{ccc}\hline
                \rule[0mm]{0mm}{4mm}Parameter & Value \rule[0mm]{0mm}{4mm}\\\hline\hline
                \rule[0mm]{0mm}{4mm}$a_1$& $9.95832\times10^{-1}$  \\
                $a_2$& $1.23972\times10^{-3}$  \\
                $b_1$& $1.19205\times10^{-2}$  \\
                $A_1$& $9.94593\times10^{-1}$  \\
                $\beta_1$& $4.93442\times10^{-1}$  \\
                $\gamma_1$& $-8.00753\times10^{-1}$  \\
                $c_1$& $-3.34000\times10^{-1}$  \\
                $d_{1}$& $2.34191\times10^{-3}$  \\
                $e_{1}$& $-1.84394\times10^{-5}$  \\
                $A_2$& $-1.13653\times10^{-1}$  \\
                $\beta_2$& $-4.10528\times10^{-1}$  \\
                $\gamma_2$& $6.79071\times10^{-1}$  \\
                $c_2$& $3.51356\times10^{-1}$  \\
                $d_{2}$& $-2.28465\times10^{-3}$  \\
                $e_{2}$& $1.80024\times10^{-5}$  \\\hline
            \end{tabular}
        \end{center}
    \end{table}
    %
 \subsection{Simulation: Asymmetric Bouc--Wen Model}
    We evaluated the performance of the control system design using the proposed method and compared it with that of a conventional method.
    The Bouc--Wen model described in (\ref{Asymmetric Bouc--Wen}) was adopted as the plant.
    The controller configuration used in the proposed method is shown in Fig. \ref{fig:MPC_based_pl}, whereas that used in the conventional method shown in Fig. \ref{fig:closed_loop}.
    The reference was sinusoidal with a width of 25 deg, an offset of 30 deg, and a frequency of 0.2 Hz.
    The sampling time was $T_{\rm s}$ = 10 ms.
    The initial PID gains were set to $\bar{\bm{\theta}}_{0} = [5.00\times 10^{-2},5.00\times 10^{-2},1.00\times 10^{-2}]^{\rm T}$, and the controller was designed using the parameter set $\bm{\theta}$ obtained via PL with E--FRIT.
    With $\lambda = 5.00\times10^{4}$, the parameters obtained through optimization were $\bm{\theta} = [1.30\times 10^{-1},1.51,6.29\times 10^{-1},7.10\times 10^{-2}]^{\rm T}$.
    In the conventional method, the controller is designed to match a PL model, including $T_{\rm c}$. Whereas, in the proposed method, the parameters of the PL model were used to design an MPC system.
    The prediction horizon $H_{\rm p}$ and control horizon $H_{\rm u}$ were set in 5 steps.
    The weight matrices are $Q(i)\equiv5I_{5}$, $R(i)\equiv0$, $V(i)\equiv I_{5}$, $\forall i$.
    Considering the hardware constraints of the experimental equipment used for the system identification, the input constraints of the MPC were set to $u_{\rm min} = 0$ V and $u_{\rm max} = 10$ V.
    %

\begin{figure}
        \centering
        \includegraphics[width=3in]{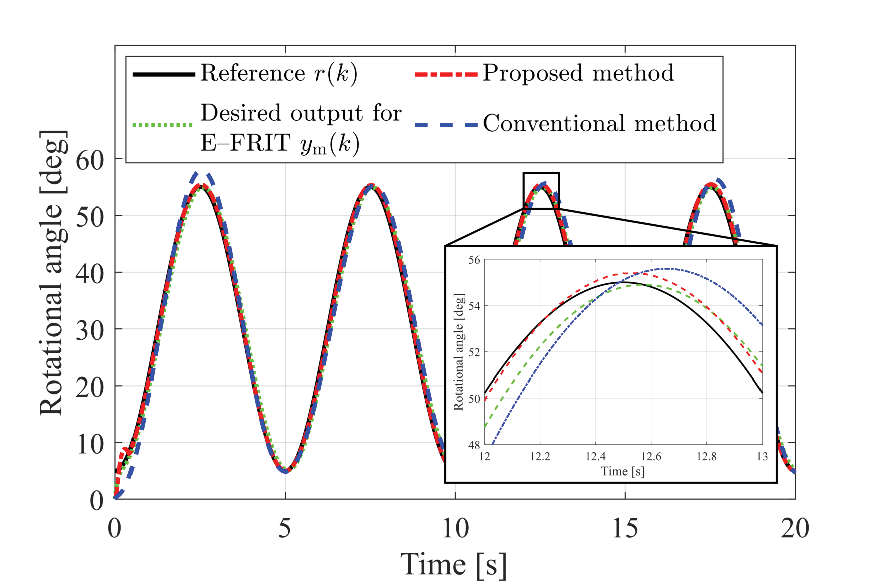}
        \caption{Comparison of tracking performance between the proposed and conventional methods in simulations of the asymmetric Bouc--Wen model.}
        \label{fig:sim_result_sin}
    \end{figure}

\begin{figure}
        \centering
        \includegraphics[width=3in]{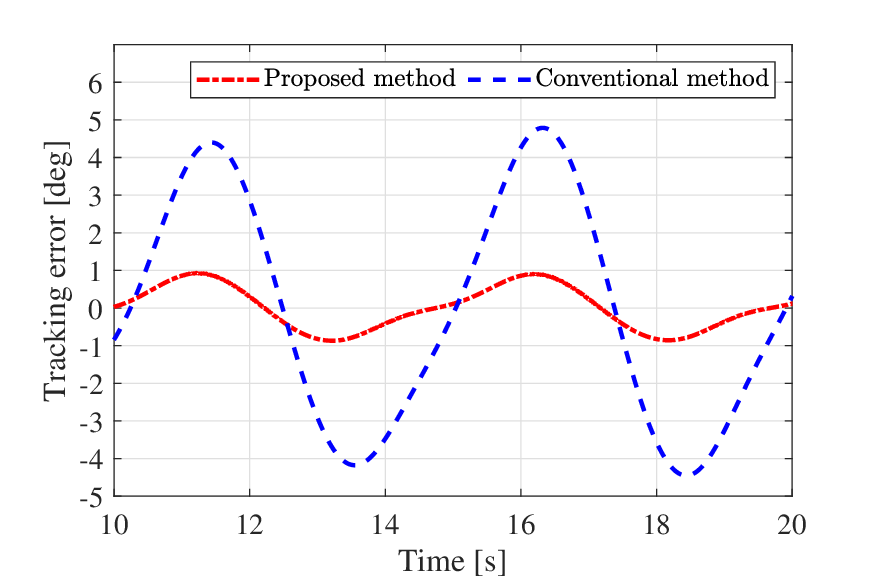}
        \caption{Comparison of tracking error in the steady--state response between the proposed and conventional methods in simulations of the asymmetric Bouc--Wen Model.}
        \label{fig:sim_error_sin}
    \end{figure}

    %
    %
    %
    The simulation results are shown in Figs. \ref{fig:sim_result_sin} and \ref{fig:sim_error_sin}.
    As shown, both design methods tracked the desired output $y_{\rm m}(k)$ generated by the reference model $G_{\rm m}(z)$. The output of the proposed method tracked the reference signal $r(k)$ with high precision.
    However, because conventional E--FRIT works to match the reference model $G_{\rm m}(z)$ and depends strongly on the given $T_{\rm c}$, it cannot track $r(k)$. By contrast, the output of the proposed controller tracks $r(k)$, hence the proposed method achieved the control objective without depending on the reference model.
    The RMSE for the proposed method is $5.65 \times 10^{-1}$ deg, whereas that of the conventional method is $2.83$ deg.
    %
    %
    %

\begin{figure}
        \centering
        \includegraphics[width=3.5in]{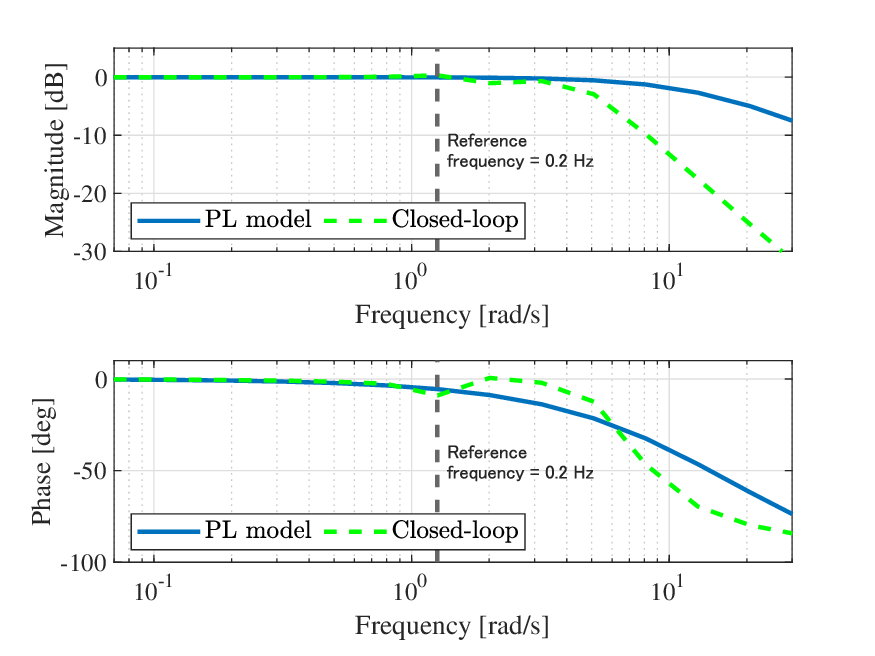}
        \caption{Comparison of Bode diagrams for closed--loop $T(\bar{\bm{\theta}}, z)$ and optimized PL model $P_{\rm L}(T_{\rm c},z)$}
        \label{fig:bode_PL}
    \end{figure}

    Next, we analyzed the model--matching performance of PL using E--FRIT.
    Previous research \cite{IFAC} indicated that the optimized PL model minimizes the matching error between the PL model and closed--loop system during the control process.
    In the present study, this was confirmed through simulations in the frequency domain.
    Specifically, using Bode plots, we visually examined the matching between the optimized PL model and tuned closed--loop system.
    The simulation conditions and the PL model parameters were identical to those used in the control simulations, as shown in Figs. \ref{fig:sim_result_sin} and \ref{fig:sim_error_sin} for the asymmetric Bouc--Wen Model.
    Fig. \ref{fig:bode_PL} presents the Bode plot of the optimized PL model obtained using $T_{\rm c} = 7.10 \times 10^{-2}$ and an optimized closed--loop system.
    From Fig. \ref{fig:bode_PL}, it is evident that, at the initial data frequency of 0.2 Hz used for pseudo--linearization, the PL model matches well with the closed--loop system.
    Moreover, the matching precision values on both sides of the frequencies (both phase and amplitude characteristics) were relatively low. They were low even for high frequencies.
    This is because, as indicated by (\ref{FRIT_evo_fc2}), the matching performed by E--FRIT is a local optimization based on the initial output data, which strongly depends on the initial reference.
    %
    %
    %
    %
    %
    %

    \section{Case Study: Tap--water--driven Artificial Muscle Control System}
    We experimentally compared the control performance of the conventional and proposed methods.
    In the experiments, we used a tap--water--driven artificial muscle actuator with one degree of freedom and an asymmetric hysteresis nonlinearity.
    \subsection{Experimental Conditions}
    The artificial muscle actuator is shown in Figs. \ref{fig:Experimental circuit} and \ref{fig:Experimental equipment} in the previous section.
    The specifications are presented in Table \ref{table:Experimental}.
    We evaluated the control performance of the proposed method and compared it with that of conventional methods for sinusoidal and square references.
    The sinusoidal reference is identical to that mentioned in Section 4.4, and the square reference $r(k)$ is given by
    \begin{equation}
        \label{ex_ref}
        r(k)=
        \left\{
        \begin{aligned}
            0, \quad &0 \leq k < 10\\
            15, \quad  &10 \leq k < 30\\
            30, \quad  &30 \leq k < 50\\
            60, \quad  &50 \leq k < 70\\
            45, \quad  &70 \leq k < 100\\
        \end{aligned}
        \right.
        .
    \end{equation}
    The sampling time was $T_{\rm s}$ = 10 ms.
    The initial PID gains were set to $\bar{\bm{\theta}}_{0} = [1.00\times 10^{-1},1.00\times 10^{-1},1.00\times 10^{-2}]^{\rm T}$, and the controller was designed using the parameter set $\bm{\theta}$ obtained via the PL with E--FRIT.
    With $\lambda = 5.00\times10^{2}$, the parameters obtained through optimization were $\bm{\theta}_{\rm sin} = [3.27 \times 10^{-1}, 5.92\times 10^{-1},9.19\times 10^{-3},1.20\times 10^{-1}]^{\rm T}$ for the sinusoidal reference and $\bm{\theta}_{\rm sq} = [1.45 \times 10^{-1}, 1.52\times 10^{-1}, 1.11\times 10^{-1}, 9.10\times 10^{-1}]^{\rm T}$ for the square reference.
    The controller in the conventional method was designed to match a PL model, including $T_{\rm c}$. On the other hand, in the proposed method, the parameters of the PL model are used to design the MPC.
    The prediction horizon $H_{\rm p}$ and control horizon $H_{\rm u}$ were set to 5.
    The weight matrices were $Q(i)\equiv4.3I_{5}$, $R(i)\equiv2.5I_{5}$, and $V(i)\equiv4I_{5}$, $\forall i$ for the sinusoidal reference and $Q(i)\equiv3I_{5}$, $R(i)\equiv0.2 I_{5}$, and $V(i)\equiv I_{5}$, $\forall i$ for the square reference.
    Considering the hardware constraints of the experimental device used for the system identification, the input constraints of the MPC were set to $u_{\rm min} = 0$ V and $u_{\rm max} = 10$ V.
    %
    %
    \subsection{Experimental Results and Discussion}
\begin{figure}
        \centering
        \includegraphics[width=3in]{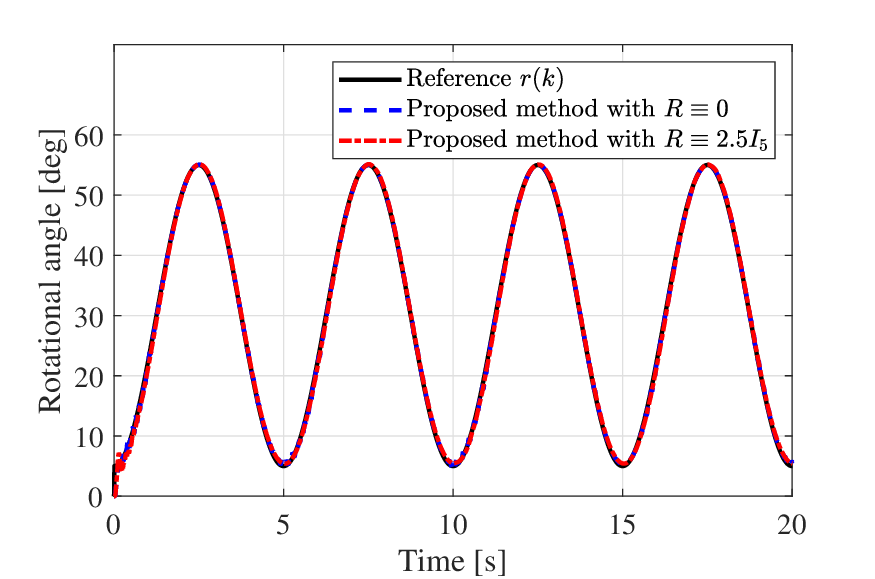}
        \caption{Comparison of tracking performance with and without input variation weights $R(i)$ in the experimental results for the sinusoidal reference.}
        \label{fig:R1}
    \end{figure}

\begin{figure}
        \centering
        \includegraphics[width=3in]{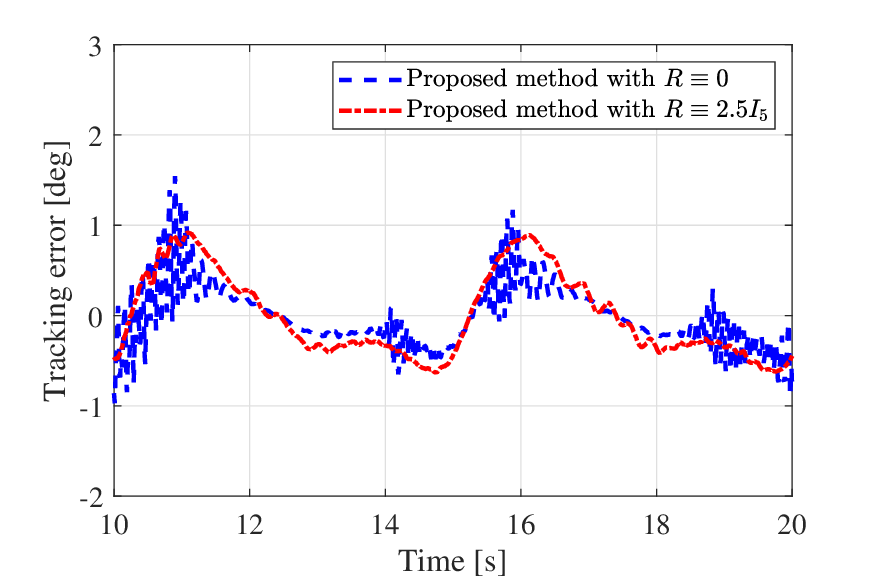}
        \caption{Comparison of the tracking errors in the steady--state responses with and without input variation weights $R(i)$ in the experimental results for the sinusoidal reference.}
        \label{fig:R2}
    \end{figure}

\begin{figure}
        \centering
        \includegraphics[width=3in]{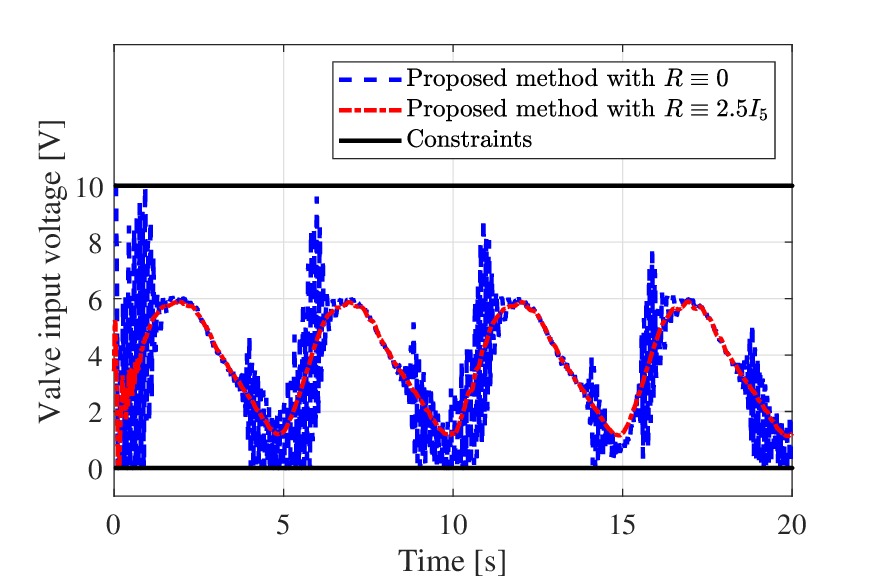}
        \caption{Comparison of the control inputs ($u(k)$) to the plant with and without input variation weights $R(i)$ in the experimental results for the sinusoidal reference.}
        \label{fig:R3}
    \end{figure}

\begin{figure}
        \centering
        \includegraphics[width=3in]{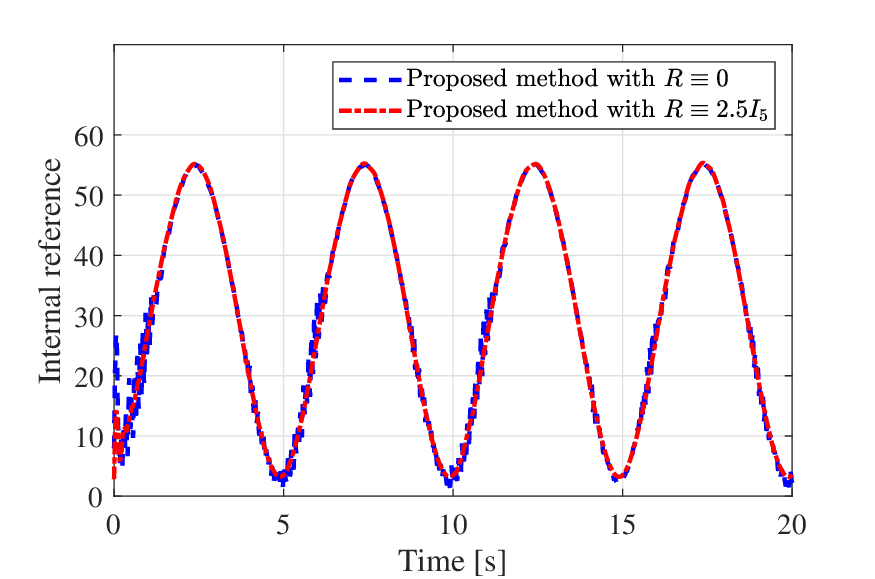}
        \caption{Comparison of the internal reference $v(k)$ with and without input variation weights $R(i)$ in the experimental results for the sinusoidal reference.}
        \label{fig:R4}
    \end{figure}

    First, we discuss the effectiveness of the second term in the cost function of MPC and its weight $R(i)$ in (\ref{MPC_evo}).
    Figures \ref{fig:R1}--\ref{fig:R4} present comparisons of the control results with and without $R(i)$ as a sinusoidal reference.
    As shown in Fig. \ref{fig:R2}, the output with $R(i) \equiv 0$ oscillated around the reference value.
    This is because the input $u(k)$ to the plant oscillates as shown in Fig. \ref{fig:R3} owing to factors such as noise.
    However, as shown in Fig. \ref{fig:R4}, the internal reference $v(k)$ did not oscillate significantly even for $R(i) \equiv 0$, indicating that the weights $V(i)$ were appropriately designed.
    Therefore, it is necessary to consider not only the internal reference $v(k)$ but also the internal input $u(k)$ to suppress oscillations in these cases.
    The experimental results indicated that in such cases, the second term was necessary, and the appropriate design of the weight $R(i)$ was effective.

\begin{figure}
        \centering
        \includegraphics[width=3in]{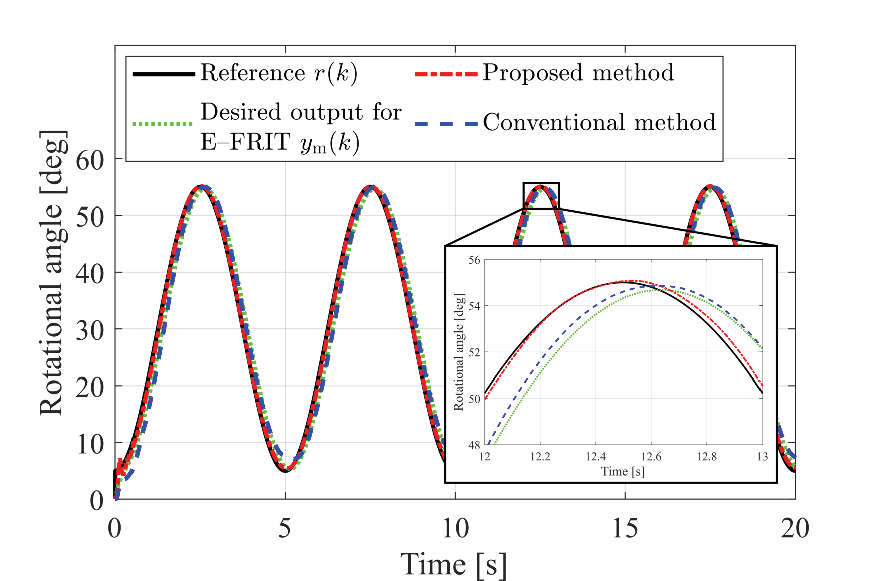}
        \caption{Comparison of tracking performance between the proposed and conventional methods in the experimental results for the sinusoidal reference.}
        \label{fig:Ex_control_result_sin}
    \end{figure}

\begin{figure}
        \centering
        \includegraphics[width=3in]{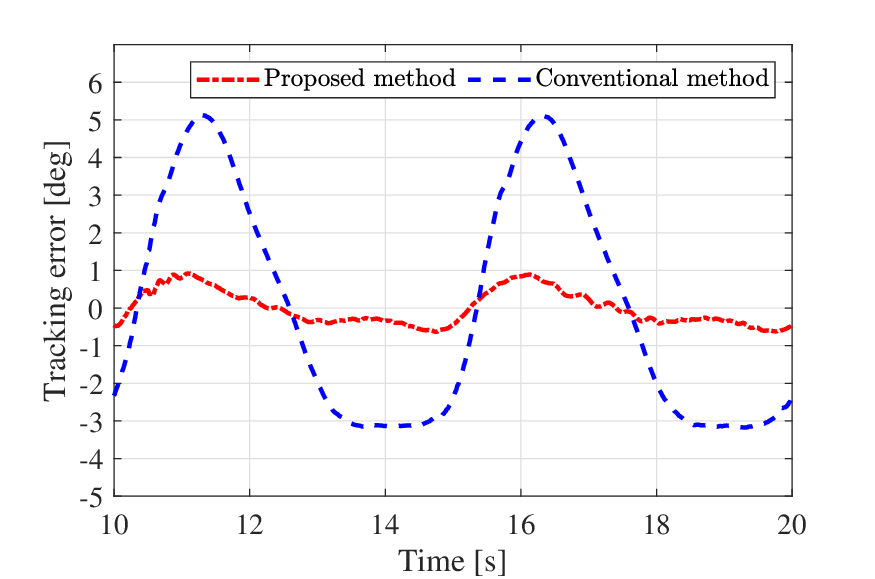}
        \caption{Comparison of tracking errors in the steady--state responses between the proposed and conventional methods in the experimental results for the sinusoidal reference.}
        \label{fig:Ex_error_sin}
    \end{figure}

    Next, we compare the control performances of the proposed and conventional methods.
    %
    %
    %
    As shown in Figs. \ref{fig:Ex_control_result_sin} and \ref{fig:Ex_error_sin}, the experimental results for the control performance were similar to the simulation results presented in Section 5.4.
    Both the design methods achieved tracking.
    The output of the conventional method tracks the desired output $y_{\rm m}(k)$ generated by the reference model $G_{\rm m}(z)$, whereas the output of the proposed method tracks the reference signal $r(k)$ with a high precision.
    Therefore, the proposed method achieved the control objective and did not depend on the reference model.
    On the other hand, as conventional E--FRIT works to match the reference model $G_{\rm m}(z)$, it tracks the desired output $y_{\rm m}(k)$, which depends more strongly on the given $T_{\rm c}$ than on tracking $r(k)$.

\begin{figure}
        \centering
        \includegraphics[width=3in]{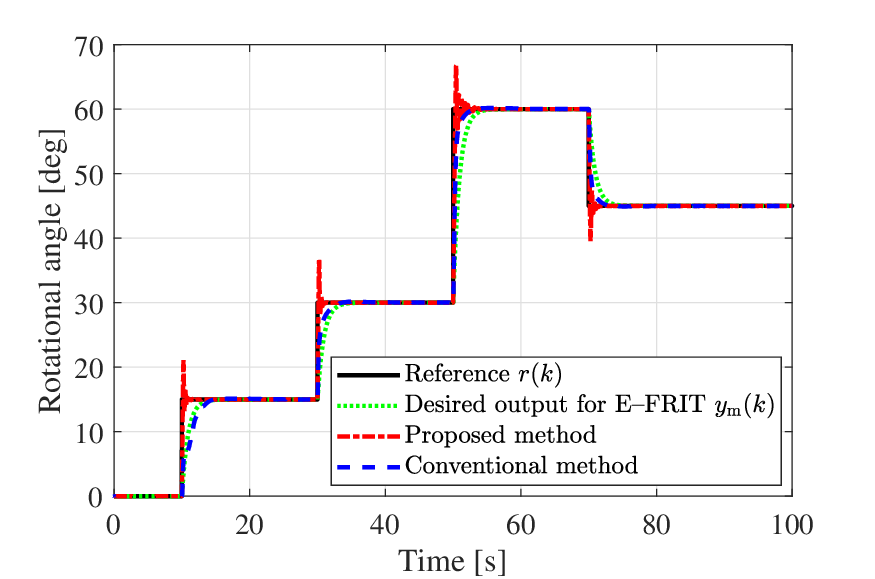}
        \caption{Comparison of tracking performance between the proposed and conventional methods in the experimental results for the square reference.}
        \label{fig:Ex_control_result_sq}
    \end{figure}

\begin{figure}
        \centering
        \includegraphics[width=3in]{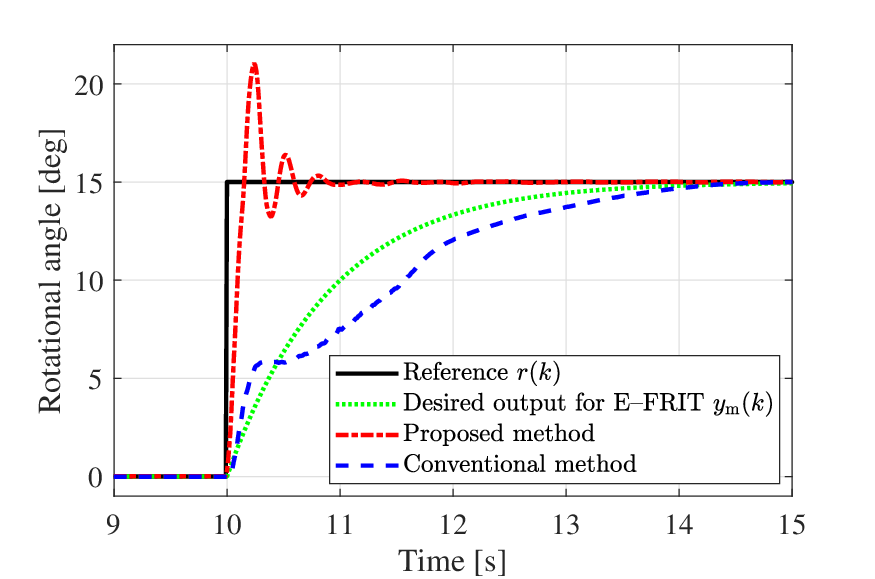}
        \caption{Comparison of tracking performance between the proposed and conventional methods in the experimental results for the square reference, which is an enlargement of Fig. \ref{fig:Ex_control_result_sq} from the initial rise from 9.5 to 15 s.}
        \label{fig:Ex_control_result_sq_enlarge}
    \end{figure}

\begin{figure}
        \centering
        \includegraphics[width=3in]{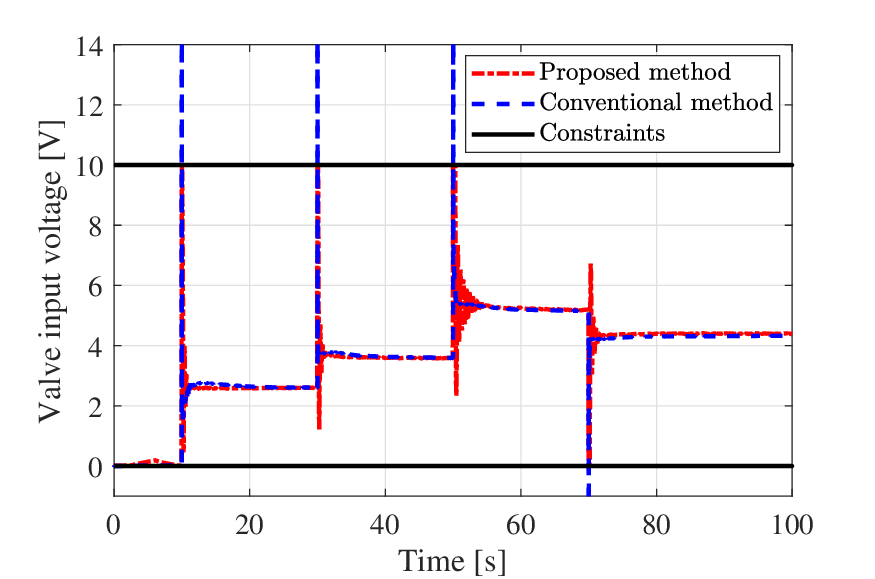}
        \caption{Control input $u(k)$ to the plant in the experimental results for the square reference.}
        \label{fig:Ex_input_sq}
    \end{figure}

    As shown in Figs. \ref{fig:Ex_control_result_sq}--\ref{fig:Ex_input_sq}, the experimental results for the control performance were similar to the simulation results presented in Section 5.1.
    In particular, the proposed method considers input constraints.
    The control performance was evaluated based on the RMSE and standard deviation (SD) from the experiment, and the results are presented in Tables \ref{table:Ex_result_sin} and \ref{table:Ex_result_sq}.
    As indicated by the quantitative evaluation results in these tables, the proposed method achieved superior control performance.
    \begin{table}
        \begin{center}
            \caption{Evaluation of control performance for the sinusoidal reference.}
            \label{table:Ex_result_sin}
            \begin{tabular}{ccc}\hline
                \rule[0mm]{0mm}{4mm}Index&Proposed method&Conventional  method\rule[0mm]{0mm}{4mm}\\\hline\hline
                \rule[0mm]{0mm}{4mm}$\rm{RMSE\ (0\ s\ to\ 100\ s)}$& $4.85 \times 10^{-1}$ deg& $2.88$ deg \\
                \rule[0mm]{0mm}{4mm}$\rm{SD\ (0\ s\ to\ 100\ s)}$& $6.77 \times 10^{-2}$ deg& $1.36 \times 10^{-2}$ deg\\\hline
            \end{tabular}
        \end{center}
    \end{table}
    \begin{table}
        \begin{center}
            \caption{Evaluation of control performance for the square reference.}
            \label{table:Ex_result_sq}
            \begin{tabular}{ccc}\hline
                \rule[0mm]{0mm}{4mm}Index&Proposed method&Conventional method\rule[0mm]{0mm}{4mm}\\\hline\hline
                \rule[0mm]{0mm}{4mm}$\rm{RMSE\ (0\ s\ to\ 100\ s)}$& $1.23$ deg& $1.98$ deg\\
                \rule[0mm]{0mm}{4mm}$\rm{SD\ (0\ s\ to\ 100\ s)}$& $3.33 \times 10^{-3}$ deg& $1.11 \times 10^{-2}$ deg\\
                \rule[0mm]{0mm}{4mm}$\rm{RMSE\ (80\ s\ to\ 100\ s)}$&$5.76 \times 10^{-3}$ deg&$2.92 \times10^{-2}$ deg\\
                \rule[0mm]{0mm}{4mm}$\rm{SD\ (80\ s\ to\ 100\ s)}$&$2.53 \times 10^{-5}$ deg&$7.16 \times 10^{-3}$ deg\\\hline
            \end{tabular}
        \end{center}
    \end{table}
    %
    %
    %
    %
    %
    %
    %
    %

    \section{Conclusion}
    In this study, we proposed a novel model predictive controller using an optimized PL model based on E--FRIT, providing a solution to the problem of control performance degradation due to the design of an improper reference model and consideration of input constraints in the E--FRIT design.

    The control performance of conventional E--FRIT strongly depended on the reference model.
    To solve this problem, we introduce a PL model and constructed a model--based controller in the outer loop.
    This allowed us to distinguish the design of the PL model from the tracking performance of the reference.
    Specifically, a model predictive controller was designed as the outer loop controller.
    In addition, the input constraints of an unknown plant were considered by estimating the plant input according to an internal reference.

    Simulations were conducted using two nonlinear classes, that is, the Hammerstein and asymmetric Bouc--Wen models. The obtained results confirmed that PL is more suitable for reference tracking control than the conventional E--FRIT.
    Furthermore, compared with conventional E--FRIT, it was shown that the control performance of the proposed method did not depend on the PL model or the parameters of the reference model.
    Moreover, we experimentally compared the proposed method with a conventional method that uses a rotational actuator comprising two artificial muscles and having a single degree of freedom.
    The proposed method achieved superior tracking control performance for the two reference types; the experimental results were similar to the simulation results.
    In addition, we confirmed that the proposed method can explicitly consider the input constraints, which are the hardware limitations of the plant.
    However, while our method demonstrates promising results when considering input constraints and enhancing the reference tracking control, addressing the robustness of the system against variations in experimental conditions and reference frequencies requires further research.

    In future, we intend to extend the proposed method to a more robust PL that considers changes in the characteristics during operation and variations in the reference frequency by incorporating adaptive fictitious reference iterative tuning\cite{A-FRIT} into our framework.
    In addition, we will explore the development of an inner--loop controller structure that explicitly considers system nonlinearity.

\bibliography{mybibfile}

\end{document}